\documentclass[preprint,aps, showpacs]{revtex4}

\usepackage{graphicx}
\usepackage{dcolumn}
\usepackage{bm}
\usepackage{color}
\usepackage{setspace}

\newcommand{\LNMO}{La$_{2}$NiMnO$_{6}$\ }

\begin{document}


\title{Paramagnetic spin pumping}

\author{Y. Shiomi$^{\, 1}$}
\author{E. Saitoh$^{\, 1,2,3,4}$}
\affiliation{$^{1}$
Institute for Materials Research, Tohoku University, Sendai 980-8577, Japan }
\affiliation{$^{2}$
WPI Advanced Institute for Materials Research, Tohoku University, Sendai 980-8577, Japan
}
\affiliation{$^{3}$
CREST, Japan Science and Technology Agency, Tokyo 102-0076, Japan
}
\affiliation{$^{4}$
Advanced Science Research Center, Japan Atomic Energy Agency, Tokai 319-1195, Japan
}

\date{\today}

\begin{abstract}
We have demonstrated spin pumping from a paramagnetic state of an insulator \LNMO into a Pt film. Single-crystalline films of \LNMO which exhibit a ferromagnetic order at $T_{C}\approx 270$ K were grown by pulsed laser deposition. The inverse spin Hall voltage induced by spin-current injection has been observed in the Pt layer not only in the ferromagnetic phase of \LNMO but also in a wide temperature range above $T_{C}$. The efficient spin pumping in the paramagnetic phase is ascribable to ferromagnetic correlation, not to ferromagnetic order.  
\end{abstract}

\maketitle

\begin{figure}[t]
\begin{center}
\includegraphics[width=14cm]{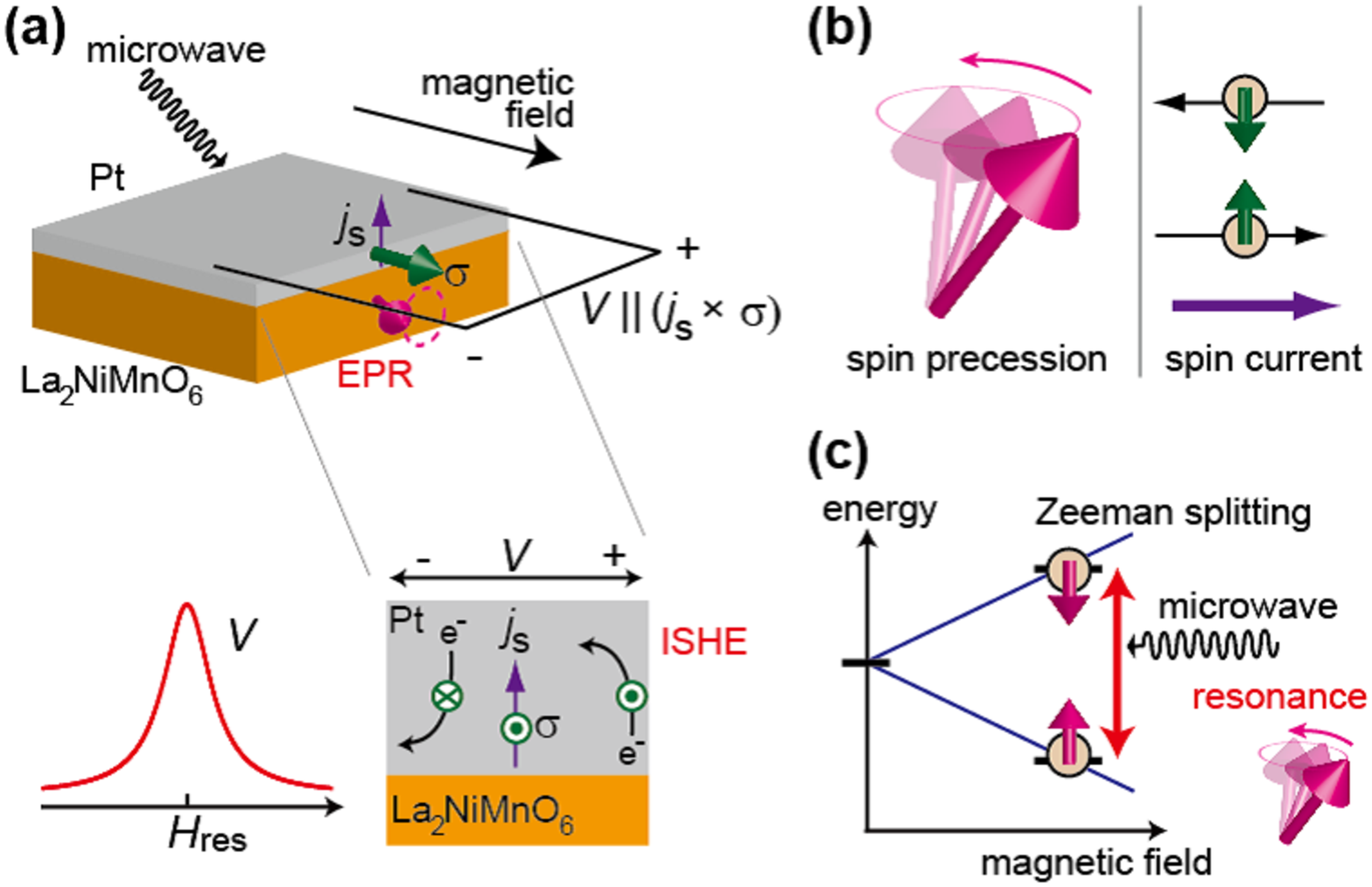}
\caption{(a) A schematic illustration of the inverse spin Hall effect induced by spin pumping in La$_{2}$NiMnO$_{6}$$\mid$Pt structures.  (b) A schematic illustration of spin pumping. Spin precession induces a pure spin current in the adjacent layer. (c) A schematic illustration of electron paramagnetic resonance (EPR). In the presence of $H$, energy levels of parallel-spin and anti-parallel-spin states are separated: the Zeeman effect. When the microwave frequency ($f$) is equal to the Zeeman energy splitting ($2\pi f = \gamma H$), electron spins precess resonantly. } 
\label{fig1}
\end{center}
\end{figure}

A spin current in nonmagnetic metals consists of electrons with opposite spins moving in opposite directions. This spin flow which does not accompany a net charge flow, called a pure spin current, is a key ingredient in spintronics devices \cite{spin-current}. A pure spin current can be detected electrically by the inverse spin Hall effect (ISHE) using metals with strong spin-orbit interaction such as Pt. In the ISHE, electrons with opposite spins are scattered into different directions by the spin-orbit interaction, as shown in Fig. \ref{fig1}(a). Then, the inverse spin Hall voltage, $V$, arises along the direction parallel to $j_{s} \times \sigma$, where $j_{s}$ and $\sigma$ are the spatial direction and the spin-polarization direction of the spin current, respectively [Fig. \ref{fig1}(a)]. \par

Spin pumping is a powerful tool to generate pure spin currents; when a magnetization precession motion is induced in a magnet {\it e.g.} by ferromagnetic resonance (FMR), a spin current is injected into an adjacent nonmagnetic metal. During the precessional motion of magnetization, the spin angular momentum in the magnet is transferred across the interface to an adjacent metal via the spin exchange at the interface, as shown in Fig. \ref{fig1}(b). The spin pumping has been studied using FMR so far in ferri- or ferro-magnet$\mid$paramagnetic-metal hybrid structures  \cite{saitoh, costache, rezende, hoffmann-prl, Ando, azevedo, silsbee, kajiwara, ohno, sandweg, castel, jungfleisch, kelly}. Pure spin currents produced by spin pumping were first observed by use of the ISHE in Ni-Fe$\mid$Pt double layers \cite{saitoh, costache, rezende, hoffmann-prl, Ando, azevedo}. Recently, the spin pumping from a ferrimagnetic insulator Y$_{3}$Fe$_{5}$O$_{12}$ to a neighboring Pt film was discovered \cite{kajiwara}. Spin pumping from magnetic insulators, which does not accompany spin-polarized electric currents across the interface \cite{ohno}, opened new interests in the spintronics field from view points of applications and also of science of spin currents \cite{sandweg, castel, jungfleisch, kelly, nakayama, flipse}.
\par

In the present study, we have developed a new insulator for spin pumping: an oxide insulator La$_2$NiMnO$_6$, whose Curie temperature, $T_{C}$, is close to room temperature. Using La$_2$NiMnO$_6$$\mid$Pt devices, we have investigated spin pumping in the temperature range from $200$ K to $350$ K. Below $T_{C}$, the spin current generated from \LNMO by FMR spin pumping is observed in an attached Pt layer via the ISHE. Furthermore, clearly, the spin-current signal appears also at {\it paramagnetic} resonance in \LNMO even far above room temperature, although \LNMO is in the paramagnetic state and it has no long-range spontaneous magnetization; the ISHE signal is observed in the Pt layer in a wide temperature range up to $350$ K. This result is the evidence of spin pumping without ferromagnetic order.  \par

\LNMO films were grown on (001) SrTiO$_{3}$ (STO) substrates by the pulsed laser deposition (PLD) technique. Polycrystalline targets were ablated with a KrF excimer laser light ($\lambda=248$ nm) with a repetition rate of $2$ Hz. During the deposition, STO substrates were kept at $800$ $^{\circ}{\rm C}$ and pure oxygen gas of $0.3$ torr was continuously supplied into the growth chamber. After the thin-film growth, the films were annealed at $800$ $^{\circ}{\rm C}$ in $400$ torr oxygen gas for $1$ hour, and then cooled down to room temperature at the rate of $15$ $^{\circ}{\rm C}$/min. The crystal structures of the films were evaluated by reflection high energy electron diffraction (RHEED) and X-ray diffraction (XRD) using Cu $K\alpha$ radiation. Clear streaks from \LNMO films were observed in RHEED patterns (see \cite{SM}). Resistivity and magnetization for the films were measured in a superconducting magnet with an electrometer and a vibrating sample magnetometer, respectively. For the ISHE measurements, a $10$-nm-thick Pt layer was sputtered in an Ar atmosphere on the top of the \LNMO films ($4\times 2$ mm$^{2}$). The simultaneous measurement of microwave resonance and dc voltage was performed using coplanar-type waveguides \cite{qiu}. The ISHE measurement at room temperature was conducted by use of an electromagnet, while that at low and high temperatures was carried out in a superconducting magnet with the incident microwave power of $1$ W due to relatively large transmission loss of incident microwave energy \cite{SM}, the small resonance absorption of \LNMO ($\sim 10$ ${\rm \mu W}$), and larger voltage noises at higher temperatures. We applied microwave by means of a signal generator or a vector network analyzer and measured resonance absorption with a microwave power meter. DC voltage generation between the ends of the Pt layer at spin resonance was detected with a nanovoltmeter. 

\begin{figure}[t]
\begin{center}
\includegraphics[width=14cm]{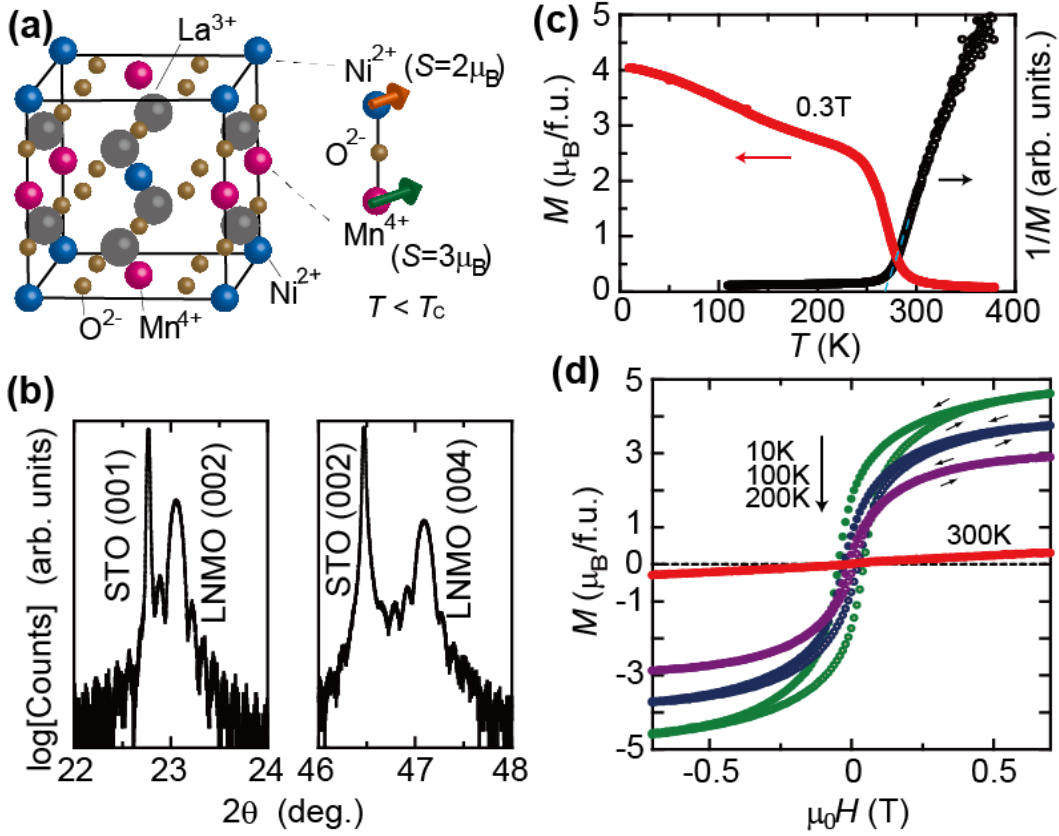}
\caption{(a) Crystal structure of ordered La$_{2}$NiMnO$_{6}$. Below the critical temperature ($T_{C}\approx 270$ K), \LNMO exhibits ferromagnetism due to superexchange interaction between Ni$^{2+}$ and Mn$^{4+}$. (b) $2\theta$-$\theta$ X-ray diffraction patterns for the $80$-nm-thick \LNMO (LNMO) film on the SrTiO$_{3}$ (STO) substrate. (c) Temperature ($T$) dependence of magnetization ($M$) at $\mu_{0}H =0.3$ T for the \LNMO film. $T$ dependence of $1/M$ is also shown in (c). $T_{C}$ is estimated to be $\sim 270$ K from the Curie-Weiss law, where the dotted line is guides to the eyes. (d) Magnetic-field ($H$) dependence of magnetization ($M$) at several temperatures. } 
\label{fig2}
\end{center}
\end{figure}

We show X-ray diffraction patterns for our samples in Fig. \ref{fig2}(b). Sharp (002) and (004) diffraction peaks of the \LNMO film were clearly observed, indicating that \LNMO films were epitaxially grown on the (001) STO substrates. The out-of-plane lattice parameter is $3.85$ \AA, which is almost the same as that of highly-ordered fillms \cite{hashisaka, kitamura}. As shown in Fig. \ref{fig2}(b), clear Laue fringes appear around (002) and (004) reflections, indicating high crystalinity, flat surface, and homogeneity of the growth film \cite{nakamura}. From the period of the fringe oscillation, thickness ($t$) of the \LNMO film was estimated to be $80$ nm.
\par

The \LNMO film was confirmed to be highly insulating; sheet resistance $R_{s}$ is higher than $20$ ${\rm M\Omega}$ below $350$ K. The high resistance at room temperature enables us to study the spin pumping free from electric conduction in La$_{2}$NiMnO$_{6}$. Temperature ($T$) and magnetic field ($H$) dependence of magnetization ($M$) is shown in Figs. \ref{fig2}(c) and \ref{fig2}(d), respectively. A clear ferromagnetic-paramagnetic transition is observed at the Curie temperature $T_{C} \sim 270$ K [Fig. \ref{fig2}(c)], which is almost the same as the bulk value \cite{rogado}. As shown in Fig. \ref{fig2}(d), the saturation magnetization at $10$ K is almost $5$ ${\rm \mu}_{B}$/f.u. in accordance with the expected value for the ferromagnetic ordering of Ni$^{2+}$ ($S=1$) and Mn$^{4+}$ ($S=3/2$) ions. Magnetic-field dependence of $M$ exhibits a ferromagnetic hysteresis behavior below $T_{C}$ (with a coercive field of $\sim 50$ mT at $10$ K). On the other hand, $M$ at $300$ K shows linear $H$-dependence, characteristic of the paramagnetic phase [Fig. \ref{fig2}(d)]. 
\par 

The ISHE measurement was performed in both ferromagnetic and paramagnetic states of La$_{2}$NiMnO$_{6}$, as illustrated in Fig. \ref{fig1}(a), where spin dynamics in the \LNMO was excited by applying a microwave. Here, spin dynamics in a ferromagnetic state is characterized by ferromagnetic resonance (FMR), while that in a paramagnetic state by electron paramagnetic resonance (EPR). FMR corresponds to a precessional motion of the order parameter, {\it i.e.} magnetization, of a ferromagnet. On the other hand, EPR is microwave absorption by individual spin excitation in an external magnetic field [Fig. \ref{fig1}(c)]; when microwave frequency is equal to the magnitude of spin-energy splitting induced by the external magnetic field, transition across the splitting is excited resonantly.
\par

\begin{figure}[t]
\begin{center}
\includegraphics[width=14cm]{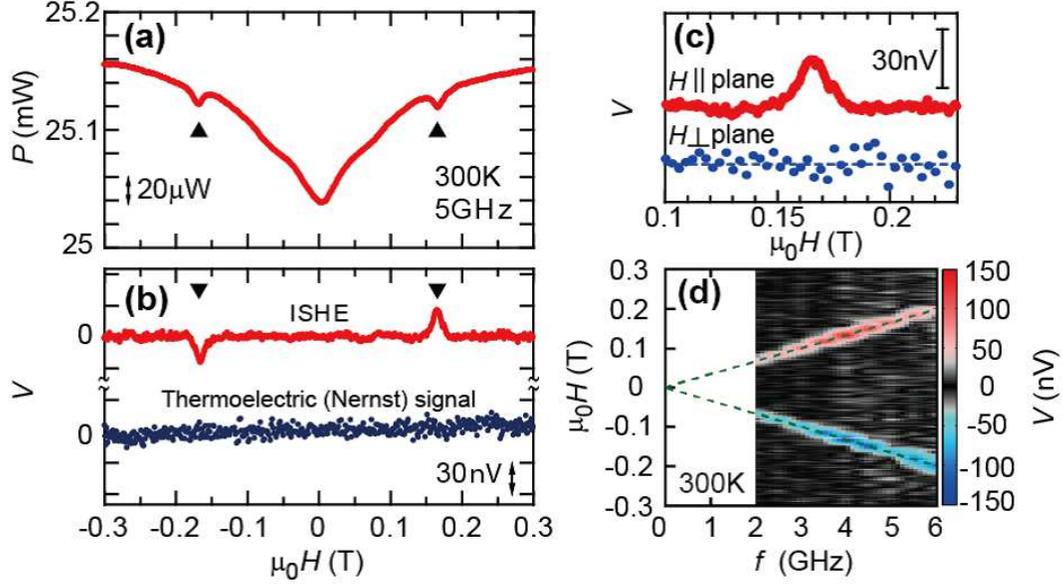}
\caption{(a),(b) Magnetic field ($H$) dependence of (a) the EPR spectrum ($P$) and (b) the inverse spin Hall voltage ($V$) for the La$_{2}$NiMnO$_{6}$$\mid$Pt($10$ nm) film. The microwave frequency is kept at $5$ GHz and the incident power is $100$ mW. The measurement temperature is $300$ K. A thermoelectric (Nernst) signal measured under a perpendicular temperature gradient (see text) is shown for comparison in (b); only a tiny normal Nernst signal (with a positive slope) is observed. (c) Comparison of voltage signal ($V$) between the cases of in-plane $H$ ($H \perp j_{s}$) and perpendicular-to-plane $H$ ($H || j_{s}$). A dotted line is a guide for the eyes. (d) A contour plot of $V$ as functions of frequency ($f$) and $H$ at $300$ K. Incident microwave power ($P_{in}$) is $0.8$ W. Dotted lines (green) are guides to the eyes.  }
\label{fig3}
\end{center}
\end{figure}

In Fig. \ref{fig3}, we show results of the spin-pumping measurement in a La$_{2}$NiMnO$_{6}$$\mid$Pt device at room temperature, where \LNMO is in the paramagnetic state. Figure \ref{fig3}(a) shows a power spectrum of the microwave reflected from the La$_{2}$NiMnO$_{6}$$\mid$Pt, $P$, as a function of magnetic field ($H$) at $300$ K. Here, incident microwave frequency ($f$) and power ($P_{in}$) are $f=5$ GHz and $P_{in}=100$ mW, respectively. The power spectrum shows dips around $\pm 0.2$ T, which correspond to microwave absorption by EPR in La$_{2}$NiMnO$_{6}$. Voltage signal in the Pt layer, $V$, measured simultaneously with the $P$ measurement is shown as a function of $H$ in Fig. \ref{fig3}(b). At this temperature, the \LNMO is in its paramagnetic phase and there is no long-range spontaneous magnetization. Nevertheless, at the field where EPR occurs, clear electromotive force peaks appear [see Fig. \ref{fig3}(b)]. The \LNMO film is a good insulator whose resistance is larger than $50$ ${\rm M\Omega}$ at room temeprature, meaning that the voltage signal is generated in the Pt layer. The sign of the voltage peak is reversed by reversing the $H$ direction and the magnitude of the peak increases linearly with microwave power \cite{SM}. As shown in Fig. \ref{fig3}(c), we also confirmed that the voltage signal disappears when $H$ is applied along a perpendicular direction to the film plane ($H || \sigma || j_{s}$). These results are consistent with the symmetry of ISHE [$V || (j_{s} \times \sigma)$].  
\par

To further confirm that the voltage signal appears in the paramagnetic state, we show, in Fig. \ref{fig3}(d), the $f$ dependence of the resonance magnetic-field, $H_{r}$, determined from the peak positions in the ISHE measurements at $300$ K. In FMR, the relation between $H_{r}$ and the resonance frequency, $f_{r}$, is given by the Kittel formula: $2\pi f_{r}=\gamma \sqrt{ \mu_{0} H_{r}(\mu_{0}H_{r}+M_{eff})} $, where $M_{eff}$ is the effective spontaneous magnetization which also includes surface anisotropy. As shown in Fig. \ref{fig3}(d), $f_{r}$ dependence of $H_{r}$ is linear at $300$ K, which indicates that $M_{eff} = 0$, confirming the EPR condition ($2\pi f_{r}=\gamma \mu_{0}H_{r}$). From the slope value, we estimated $\gamma$ to be $\gamma = 1.88 \times 10^{11}$ ${\rm T}^{-1}{\rm s}^{-1}$, which is almost the same as the free electron's gyromagnetic ratio ($\gamma_{e} = 1.76 \times 10^{11}$ ${\rm T}^{-1}{\rm s}^{-1}$). This result demonstrates again the spin-current injection from the paramagnetic state.
\par

The observed peak voltage ($V_{0}$) at EPR is not due to heating effects induced by microwave resonance absorption. To examine possible contamination of thermoelectric voltage in the Pt layer, we show, in Fig. \ref{fig3}(b), the Nernst-effect signal induced by a temperature gradient externally applied perpendicularly to the film plane. Here, temperature difference was applied with the use of a $1$k${\rm \Omega}$-resistance heater attached on the STO substrate and the applied power $= 0.4$ W was set to almost the same as the incident microwave power in the ISHE measurement ($P_{in}=0.1$-$1$ W). As shown in Fig. \ref{fig3}(b), the observed thermoelectric signal is proportional to $H$ and its magnitude is very small; this voltage corresponds to the ordinary Nernst effect (see also \cite{SM}). Because the magnitude of the observed electromotive force signal at EPR is much greater than the thermoelectric voltage despite very small resonance-absorption power $\sim 10$ ${\rm \mu W}$, we can safely rule out possible heating-induced effects in the ISHE measurements. The spin-pumping voltage normalized by the resonance absorption, which indicates the efficiency of paramagnetic spin pumping, is $\sim 30$ nV$/10$ ${\rm \mu W} = 0.003$ ${\rm V/W}$ at $5$ GHz [Figs. \ref{fig3}(a) and \ref{fig3}(b)], which is comparable to that for the FMR spin pumping in Y$_{3}$Fe$_{5}$O$_{12}$$\mid$Pt bilayers \cite{jungfleisch}. 
\par

\begin{figure}[t]
\begin{center}
\includegraphics[width=14cm]{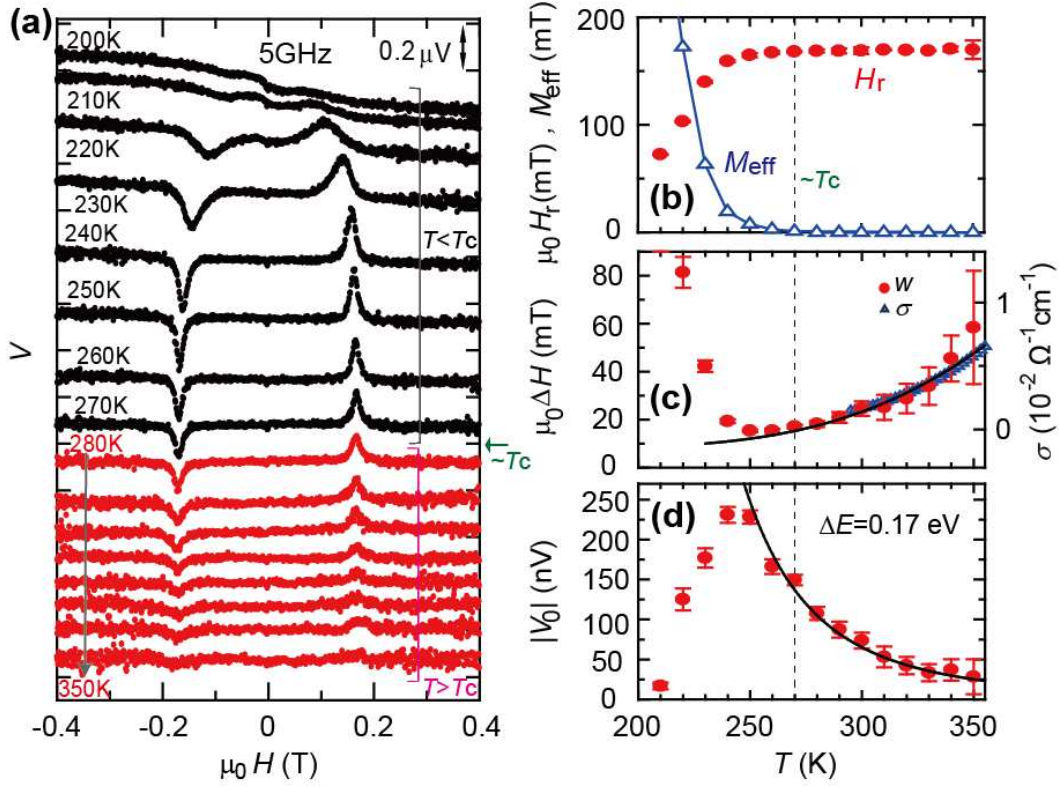}
\caption{(a) Magnetic field ($H$) dependence of inverse spin Hall voltage ($V$) at various temperatures. The microwave frequency ($f$) is $5$ GHz. (b) Temperature ($T$) dependence of the resonance field ($H_{r}$) and effective spontaneous magnetization ($M_{eff}$). The $M_{eff}$ values are estimated using the Kittel formula. (c),(d) Temperature ($T$) dependence of (c) the full-width at half-maximum ($\Delta H$) and (d) the magnitude ($|V_{0}|$) of the peaks in the inverse spin Hall voltage signals. The solid curves in (c) and (d) are fits to Arrhenius equations. Temperature ($T$) dependence of electrical conductivity ($\sigma$) for the \LNMO film is shown for comparison in (c). The dotted vertical lines in (b)-(d) indicate $T_{C}$ ($\approx 270$ K). }
\label{fig4}
\end{center}
\end{figure}

We show temperature ($T$) dependence of the voltage signals between $200$ K and $350$ K, in Fig. \ref{fig4}(a). Voltage peaks which correspond to the ISHE induced by spin pumping are observed at each $T$; it is also noted that small spin Seebeck voltage signals \cite{uchida} are observed at $200$ K and $210$ K as a broad background, which is asymmetric with respect to $H$, because of the heating in the waveguide due to microwave energy loss (see also \cite{SM}). Below $T_{C} \approx 270$ K, the resonance field, or the voltage-peak position $H_{r}$ significantly changes with $T$, while $H_{r}$ is almost constant above $T_{C}$. In Fig. \ref{fig4}(b), $M_{eff}$ values are estimated from the $H_{r}$ values using the Kittel formula at each $T$. As $T$ increases from $200$ K, the $M_{eff}$ value rapidly decreases and becomes zero above $T_{C}$. The temperature dependence of $M_{eff}$ is similar to that of the observed magnetization ($M$) at $0.3$ T shown in Fig. \ref{fig2} (c). The result confirms again that the \LNMO is in its paramagnetic ($M_{eff}=0$) phase at $300$ K where the ISHE signal was clearly observed. 
\par

A short-range ferromagnetic correlation above $T_{C}$, which was observed in La$_{2}$NiMnO$_{6}$ \cite{raman, xmcd, esr, zhou}, can be responsible for the persistent spin-pumping signals in such a wide $T$-range above $T_{C}$ in La$_{2}$NiMnO$_{6}$$\mid$Pt; the short-range correlation was observed above $T_{C}$ up to $350$-$400$ K, {\it e.g.} by Raman spectroscopy \cite{raman}, and also manifests itself as small deviation of the inverse susceptibility from the Curie-Weiss behavior above $T_{C}$ [Fig. \ref{fig2}(c)]. The spin-pumping amplitude is proportional to the two-point correlation function for local magnetization \cite{sp-jpsj}, while the static magnetization is proportional to local magnetization itself. Thanks to the short-range correlation, the spin pumping signal comparable to the ferromagnetic spin pumping is observed even in the paramagnetic state of La$_{2}$NiMnO$_{6}$. Though such a short-range correlation has been observed also in other ferromagnets above $T_{C}$ \cite{dms1, lsco}, the high $T_{C}$ and strong short-range correlation of \LNMO enable the efficient paramagnetic spin pumping at room temperature.
\par

We fit the ISHE voltage signals around microwave resonance (EPR or FMR) using Lorentz functions and plot the obtained half-maximum full-width (HMFW, $\Delta H$) and peak magnitude $|V_{0}|$ as a function of $T$ in Figs. \ref{fig4}(c) and \ref{fig4}(d), respectively. As $T$ increases from $200$ K, $\Delta H$ shows a minimum at $250$ K, and then increases gradually above $T_{C}$ [Fig. \ref{fig4}(c)]. On the other hand, the peak voltage $|V_{0}|$ increases with increasing $T$ from $200$ K and shows a maximum at $250$ K [Fig. \ref{fig4}(d)]. Above $250$ K, $|V_{0}|$ decreases with increasing $T$, but it is still prominent even far above $T_{C}$ up to $350$ K [highlighted in red in Fig. \ref{fig4}(a)]. As shown in Figs. \ref{fig4}(c) and \ref{fig4}(d), the overall $T$ dependence of $|V_{0}|$ is similar to that of the inverse of the linewidth, $1/\Delta H$; the observed enhancement of $|V_{0}|$ around $T_{C}$ can be attributed to the suppression of $\Delta H$. 
\par

The linewidth $\Delta H$, which reflects the damping for spin precession, exhibits Arrhenius-type $T$ dependence above $T_{C}$, as shown in Fig. \ref{fig4}(c): $\Delta H \propto \exp[ -\Delta E/ k_{B}T ]$. Here, the activation energy $\Delta E$ for $\Delta H$ is $0.17$ eV. We found that this $\Delta E$ value is the same as that for the electrical conductivity of the \LNMO film, $\sigma$ [$= 1/(R_{s}t)$], as shown in Fig. \ref{fig4}(c). The conductivity-like $T$-dependence of $\Delta H$ implies that spin relaxation occurs via thermally excited carriers above $T_{C}$. The exponential $T$ dependence of $|V_{0}|$ $(\propto 1/\Delta H)$ above $T_{C}$ [Fig. \ref{fig4}(d)] shows that the paramagnetic spin pumping is irrelevant to spin waves possibly subsisting above $T_{C}$, {\it e.g.} paramagnons, which should be related with the $T$ dependence of field-induced magnetization.
\par

As for the lower-$T$ behavior of $\Delta H$ ($\sim 1/V_{0}$), we note that such a minimal resonance linewidth and a maximal resonance absorption near $T_{C}$ as observed in Figs. \ref{fig4}(c) and \ref{fig4}(d) have been reported in doped perovskite manganese oxides $A_{1-x}A^{\prime}_{x}$MnO$_{3}$ ($A=$La, Pr, ...; $A^{\prime}=$Ca, Sr, Ba,...) \cite{mn-adv}. For La$_{0.67}$Ca$_{0.33}$MnO$_3$ \cite{oseroff} and La$_{0.67}$Sr$_{0.33}$MnO$_3$ \cite{causa}, as $T$ approached $T_{C}$ from above, the resonance linewidth was reported to go through a minimum and then to increase abruptly below $T_{C}$; this is similar to the present case [Fig. \ref{fig4}(c)]. In the previous literature \cite{mn-adv, ramos}, the increase in the linewidth below $T_{C}$ has been attributed to intrinsic magnetic inhomogeneity \cite{mn-adv} or a nonuniform demagnetization tensor \cite{ramos}. Similar effects may cause the increase in $\Delta H$ below $T_{C}$ for the present \LNMO samples.   
\par

In summary, we have demonstrated spin pumping above the Curie temperature in a double-perovskite oxide insulator La$_{2}$NiMnO$_{6}$. The paramagnetic spin pumping in La$_{2}$NiMnO$_{6}$$\mid$Pt which we found operates at room temperature and, in spite of the absence of long-range magnetic order, its efficiency can be comparable to that of typical spin-pumping devices including ferromagnets. Since spin mixing conductance which has been believed to be the language of spin pumping cannot be defined for the paramagnetic state, the paramagnetic spin pumping will require expansion of the concept of spin pumping. 
\par

We thank M. Kawasaki, A. Sawa, and Y. Kajiwara for experimental help and H. Adachi, D. Hou, R. Iguchi, K. Uchida, and Y. Fujikawa for fruitful discussions. This work was supported by CREST-JST ``Creation of Nanosystems with Novel Functions through Process Integration", Grant-in-Aid for Scienctific Research (A) (24244051) from MEXT, the Murata Science Foundation, and the Nippon Sheet Glass Foundation for Materials Science and Engineering.

\newpage

\newpage

\newpage

\newpage

\clearpage

\section*{{\large Supplemental Material for ``Paramagnetic spin pumping" }}

\begin{center}
{\large Y. Shiomi and E. Saitoh}  \par
\end{center}

%


\section*{S1. RHEED pattern for the \LNMO thin film}

\begin{figure}[h]
\begin{center}
\includegraphics[width=12cm]{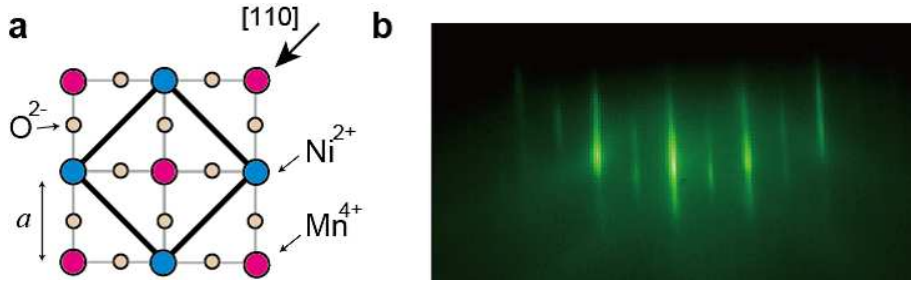}
\caption{(Fig. S1)({\bf a}) A schematic illustration of the crystal structure of ordered La$_{2}$NiMnO$_{6}$. ({\bf b}) Reflection high energy electron diffraction (RHEED) pattern for the \LNMO thin film with a $[110]$ beam incidence. Clear twofold streaks are observed.  } 
\label{figS1}
\end{center}
\end{figure}

Reflection high energy electron diffraction (RHEED) is a useful method to confirm the ordering of $B$ site ions in the double-perovskite ($A_{2}B^{\prime}B^{\prime \prime}$O$_{6}$) structure [M. Hashisaka, {\it et al.} Appl. Phys. Lett. {\bf 89}, 032504 (2006)]. In Fig. \ref{figS1}, we show a RHEED pattern for the \LNMO film with a $[110]$ beam incidence. In the ordered double-perovskite structure, $B^{\prime}$ and $B^{\prime \prime}$ cations are alternately arranged at the $B$ sites [Figs. 2(a) and \ref{figS1}(a)]. As shown in Fig. \ref{figS1}(b), clear streaks from a twofold superstructure are observed for the \LNMO film, which shows the ordering of Mn$^{4+}$ and Ni$^{2+}$ ions in the $B$ sites. 
\par

\newpage

\section*{S2. Spin Seebeck effect for the La$_{2}$NiMnO$_{6}$$\mid$Pt films}

\begin{figure}[h]
\begin{center}
\includegraphics[width=10cm]{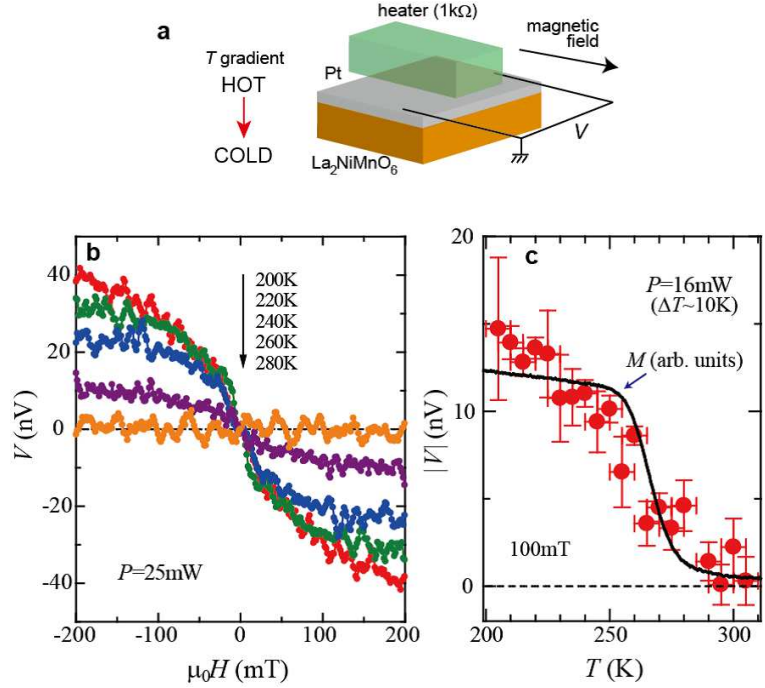}
\caption{(Fig. S2)({\bf a}) A schematic illustration of the experimental setup for the spin Seebeck effect. Temperature difference in the perpendicular direction is applied using a chip resistance ($1$ k${\rm \Omega}$) heater attached on the Pt layer. ({\bf b}) Magnetic-field ($H$) dependence of spin Seebeck voltage ($V$) at some temperatures in the La$_{2}$NiMnO$_{6}$$\mid$Pt film. The applied power to the heater ($P$) is $25$ mW. ({\bf c}) Temperature ($T$) dependence of spin Seebeck voltage ($V$) at $100$ mT. The temperature difference ($\Delta T$) is roughly estimated to be $10$ K from the change in resistance of the Pt layer. Magnetization ($M$) at $100$ mT is shown for comparison (black curve).  } 
\label{figS2}
\end{center}
\end{figure}

In ferromagnet$\mid$paramagnetic-metal bilayers, ISHE voltage is generated in a paramagnetic metal as a result of a perpendicular temperature gradient  (the longitudinal Seebeck effect) [K. Uchida, {\it et al.} Appl. Phys. Lett. {\bf 97}, 172505 (2010).]. We have investigated the longitudinal spin Seebeck effect for the La$_{2}$NiMnO$_{6}$$\mid$Pt bilayer films, as shown in Fig. \ref{figS2}. Here, a $1$k${\rm \Omega}$-resistance heater is attached on the Pt layer to generate a temperature gradient along the perpendicular direction. As shown in Fig. \ref{figS2}(b), a negative spin Seebeck signal is observed at $200$ K and it becomes smaller in magnitude monotonically with increasing system temperature. The magnitude of voltage signal ($|V|$) at $100$ mT is shown as a function of temperature ($T$) in Fig. \ref{figS2}(c). The temperature dependence of $|V|$ is almost proportional to that of magnetization ($M$) measured at $100$ mT (black curve). We note that estimation of ``paramagnetic" spin Seebeck contribution is difficult at room temperature, because the magnitude of voltage signal is very small at room temperature [Figs. 3(b) and \ref{figS2}(b)] and because  paramagnetic spin Seebeck voltage should be linear with the magnetic field similarly to the ordinary Nernst signal in Pt. \par

As shown in Fig. 4(a), in the spin-pumping measurements at $200$ K and $210$ K, small spin-Seebeck voltage is observed as a background in addition to the spin-pumping signals. The sign of the spin Seebeck voltage is negative in a positive magnetic field, which indicates that the Pt side is warmed in the spin-pumping measurements. Hence, the origin of the spin Seebeck signal is not a temperature increase in \LNMO caused by the microwave resonance absorption, but is ascribable to that in the waveguide or Pt due to microwave energy loss. It is noted that possible temperature elevation in the La$_{2}$NiMnO$_{6}$$\mid$Pt film during the spin-pumping measurements does not affect our conclusion, because ferromagnetic correlation becomes weaker at higher temperatures.   
\par

\clearpage

\section*{S3. $S_{11}$ parameters for the coplanar waveguides}
\begin{figure}[h]
\begin{center}
\includegraphics[width=7cm]{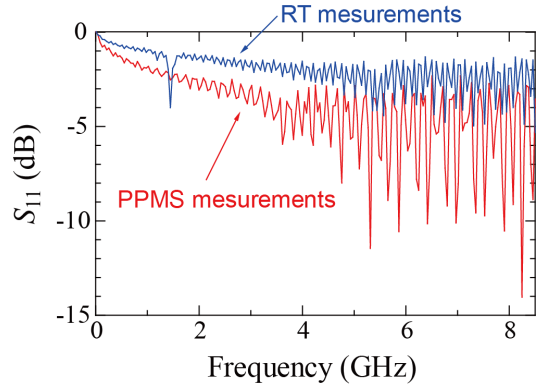}
\caption{(Fig. S3) $S_{11}$ parameters for the coplanar waveguides in room-temperature (RT) measurements and low- and high-temperature measurements in PPMS. } 
\label{figS3}
\end{center}
\end{figure}

In Fig. \ref{figS3}, we compare $S_{11}$ parameters for the reflection-type (short-end) coplanar waveguides used in the present study [the same design as that in Z. Qiu, {\it et al.}, Appl. Phys. Lett. {\bf 103}, 182404 (2013)] between room-temperature (RT) measurements and low- and high-temperature measurements in PPMS. Since the microwave cable is longer in PPMS measurements than in RT ones, the microwave loss is larger.  

\clearpage

\section*{S4. Power dependence of the inverse spin Hall voltage induced by spin pumping at room temperature}

\begin{figure}[h]
\begin{center}
\includegraphics[width=12cm]{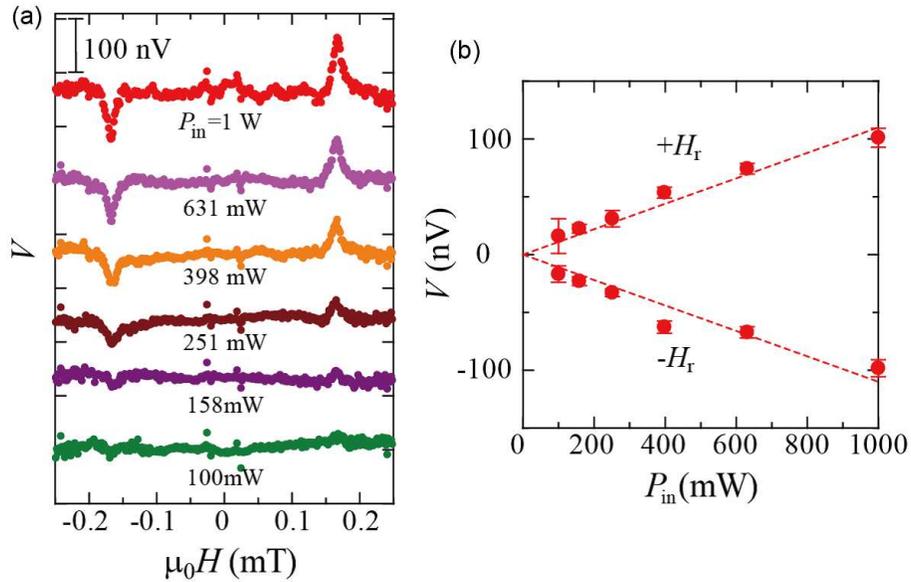}
\caption{(Fig. S4) (a) Magnetic-field ($H$) dependence of the inverse spin Hall voltage ($V$) induced by spin pumping at room temperature at different microwave power levels ($P_{in}$). The data are vertically shifted for clarity. (b) Power ($P_{in}$) dependence of the inverse spin Hall voltage induced by spin pumping at room temperature.
 } 
\label{figS4}
\end{center}
\end{figure}

In Fig. \ref{figS4}(a), we show magnetic-field ($H$) dependence of the inverse spin Hall voltage ($V$) induced by spin pumping at different microwave power levels ($P_{in}=100$ mW to $1$ W). Here, the measurement was done at room temperature and microwave frequency is $5$ GHz. With increasing $P_{in}$, the magnitude of the peak voltage increases monotonically. The values of voltage peaks are plotted as a function of $P_{in}$ in Fig. \ref{figS4}(b). The power dependence is almost linear, which is the same as the cases in ferromagnetic spin pumping.

\clearpage

\section*{S5. Analysis on resonance linewidth ($\Delta H$) at room temperature}

\begin{figure}[b]
\begin{center}
\includegraphics[width=12cm]{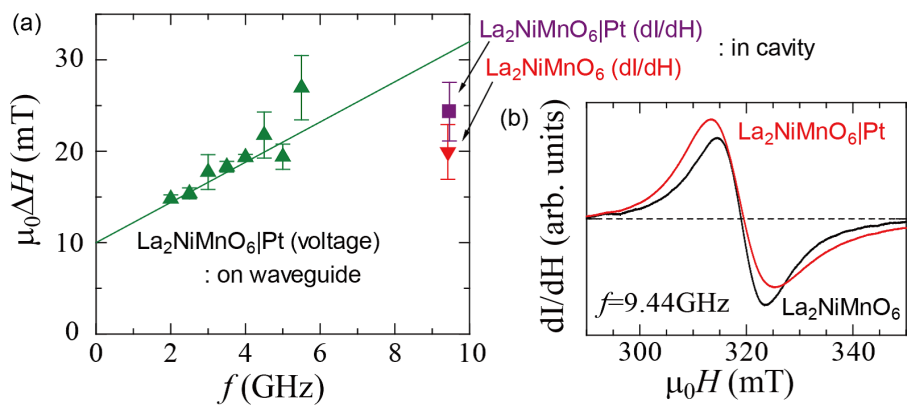}
\caption{(Fig. S5) (a) Frequency dependence of the linewidth ($\Delta H$) at room temperature. The data represented as green triangles were obtained from Lorentzian fits to the ISHE voltage in Fig. 3(d). (b) Resonance spectra for \LNMO and La$_{2}$NiMnO$_{6}$$\mid$Pt. The measurement was performed in a cylindrical cavity and microwave power was set to be $10$ mW.
 } 
\label{figS5}
\end{center}
\end{figure}

In Fig. \ref{figS5}, we show the resonance linewidth for \LNMO and La$_{2}$NiMnO$_{6}$$\mid$Pt which was measured on the colanar waveguides and using a cylindrical $9.44$-GHz TE$_{011}$ cavity. Figure \ref{figS5}(b) shows EPR spectra for \LNMO and La$_{2}$NiMnO$_{6}$$\mid$Pt measured in the cavity. Clear resonance spectra are observed around $320$ mT. In measurements for several sets of samples, the magnitude of the full-width at half-maximum ($\Delta H= \sqrt{3} \Delta H_{pp}$) is slightly scattered around $20$ mT for both \LNMO and La$_{2}$NiMnO$_{6}$$\mid$Pt as shown in Fig. \ref{figS5}(a), and precise estimation of the possible linewidth enhancement by Pt coating was difficult. \par

Figure \ref{figS5}(a) shows frequency dependence of the full-width at half-maximum ($\Delta H$) for La$_{2}$NiMnO$_{6}$$\mid$Pt measured on the waveguides at room temperature; the green triangles indicate the $\Delta H$ values obtained from Lorentzian fits to the ISHE voltage shown in Fig. 3(d). $\Delta H$ increases almost linearly with frequency. At $5$ GHz, almost half of $\Delta H$ originates in extrinsic origins independent of microwave frequency. From the value of the slope, we obtain $\alpha = \gamma \mu_{0} \Delta H / (2 \pi f) \approx 0.06$ for La$_{2}$NiMnO$_{6}$$\mid$Pt.
\par

On the temperature dependence of $\Delta H$, it is notable that according to a former study on EPR for polycrystalline \LNMO [S. Zhou, {\it et al.} Appl. Phys. Lett. {\bf 91}, 172505 (2007)], temperature dependence of $\Delta H$ for \LNMO above $T_{C}$ is similar to that of $\Delta H$ obtained from ISHE voltage in La$_{2}$NiMnO$_{6}$$\mid$Pt [$\Delta H \propto \exp[-\Delta E/k_{B}T]$ in Fig. 4(c)]. This result suggests that the possible enhancement of the linewidth by Pt coating changes with temperature in proportion to $\exp[-\Delta E/k_{B}T]$, if the linewidth enhancement is present also for paramagnetic spin pumping.
\par

\end{document}